# REQUIREMENT ANALYSIS, ARCHITECTURAL DESIGN AND FORMAL VERIFICATION OF A MULTI-AGENT BASED UNIVERSITY INFORMATION MANAGEMENT SYSTEM

Nadeem AKHTAR[1]   Aisha Shafique GHORI[1]   Nadeem SALAMAT[2]

[1]Department of Computer Science and Information Technology,
The Islamia University of Bahawalpur, Pakistan
nadeem.akhtar@iub.edu.pk

[2]Karakoram International University
15100-Gilgit-Biltastan, Pakistan

## ABSTRACT

*This paper presents an approach based on the analysis, design, and formal verification of a multi-agent based university Information Management System (IMS). University IMS accesses information, creates reports and facilitates teachers as well as students. An orchestrator agent manages the coordination between all agents. It also manages the database connectivity for the whole system. The proposed IMS is based on BDI agent architecture, which models the system based on belief, desire, and intentions. The correctness properties of safety and liveness are specified by First-order predicate logic.*

## KEYWORDS

*Information Management System (IMS), Multi-agent system, Architectural design, Formal verification, BDI (Belief, Desire, Intention) agent model, First-order predicate logic*

## 1. INTRODUCTION

An approach has been proposed for the analysis, architectural design and formal verification of an Information Management System (IMS). A multi-agent based architecture is suitable for such a system. In a multi-agent system applications are designed in terms of autonomous software entities called agents that flexibly achieve their objectives by interacting with one another in terms of high level protocols and languages [Zambonelli, Jennings and Wooldridge, 2003]. An Agent is a self-contained program capable of controlling its own decision-making, based on its perception of environment, in pursuit of one or more objectives [Jennings and Wooldridge, 1996].

A method for the architecture and formal verification of university IMS has been proposed. The Belief, Desire, Intention (BDI) agent model [Bratman, 1987] has been adopted. Each agent is autonomous and can make decisions based on its knowledge-base. An agent based on BDI theory can adapt to changing situations by focusing on the most appropriate goal at the time [Rens, Ferrein and Van, 2009].

Our objectives are to propose an architectural design of the system; formally specify and verify the correctness properties of the system; and then based on this formal foundation implement the system using an agent-based programming framework. Major agent programming platforms are; JACK [Howden et al., 2001] [Busetta et al., 1999], AgentSpeak (L) [Rao, 1996] and 3APL [Hindriks et al., 1999]. Section-2 presents problem statement; section-3 presents objectives; section-4 state of the art; section-5 multi-agent IMS; and section-6 conclusion and future work.

## 2. PROBLEM STATEMENT

The requirement analysis, architectural design, formal verification of a multi-agent based university IMS. There is a need of a correct information management, processing, and report generation system for the students and faculty of Department of Computer Science & IT, The Islamia University of Bahawalpur. Multi-agent based system provides a distributed problem solving system that has a sophisticated pattern of interactions; and using a formal base ensures system correctness.

## 3. OBJECTIVE

The *first* and foremost objective is the formal verification of correctness properties of safety and liveness. This formal verification provides a mathematically correct foundation for the architectural design and implementation.

The *second* objective is to provide the users with a state of the art platform-independent agent based IMS. An agent based system which works on computers as well as on mobile devices; provides students with information regarding their class schedules, examination date sheets, results, fee submission dates, university bus timings and routes.

The *third* objective is the automated creation of statistical graphical reports. These reports would get a measurement of the quality of education, and therefore would greatly help to uplift the quality of education.

The *fourth* objective is to propose a system that is exactly an image of human mental attitude. For this purpose BDI agent architecture is used. In BDI the belief, desire, and intention are the elements showing mental states.

## 4. STATE OF THE ART

Agent technology is one of the most promising technologies for distributed and complex systems. Software agents are entities that can make autonomous decision. The analysis, design and development of intelligent software agents touch both the artificial intelligence and software engineering areas [Geylani, 2013]. Agent characteristics include reactivity, autonomy, co-operation, and reasoning ability. A multi-agent system is a loosely coupled network of problem solvers that work together to solve problems that are beyond the individual capabilities or knowledge of each problem solver. [Demazeau, 2006] consider four essential building blocks of agent system: agent (the processing entities), interactions (the elements for structuring internal interaction between entities), organization (the elements for structuring sets of entities within the multi-agent system), and finally the environment that is defined as the domain-dependent elements for structuring external interactions between entities. Agents coordinate and cooperate with each other to accomplish tasks.

### 4.1. BDI agent architecture

The BDI agent architecture is based on the philosophical work of [Bratman, 1987], the theoretical and practical work of [Rao and Georgeff, 1995] [Jennings and Wooldridge, 1994] [Burkhard, 1996]. BDI agent architecture describes agent behaviors [Georgeff and Lanasky, 1987] based on the theory of action in humans proposed by [Bratman, 1987]. BDI is adopted to model the system and each agent works like mental attitude of human and agents can be reactive or pro-active depending on their situation [Evertsz, 2008]. Goals are first decided and then work is done on requirements to extract those goals. Belief represents the state of the real world, such as variables, database or different symbols; intentions are committed plans and procedures. Computationally, intentions may be a set of executing threads in a process that achieve the goals (desires) of the system.

### 4.2. Correctness – Safety and Liveness

Correctness properties of liveness and safety complement each other. Safety property ensures "something bad that does never happen" i.e. the program will never produce wrong result. Liveness means "a good thing happens" during execution.

## 5. MULTI-AGENT IMS: A CASE STUDY

The IMS has multiple agents that interact with other agents within the system to accomplish their tasks and achieve goals. This model allows the system extensibility. The system needs extension in a case when the system has to create a new agent in the system. Extension means to enhance some new features by making local changes in the code to maintain backward compatibility. Updating beliefs and deliberation requires little changes in the model.

### 5.1. Architecture

The proposed IMS is distributed, composed of agents. Agents are the basic building blocks of the system that receives process, manipulate and store information. The functional agents corporate and coordinates with each other to accomplish system functionality.

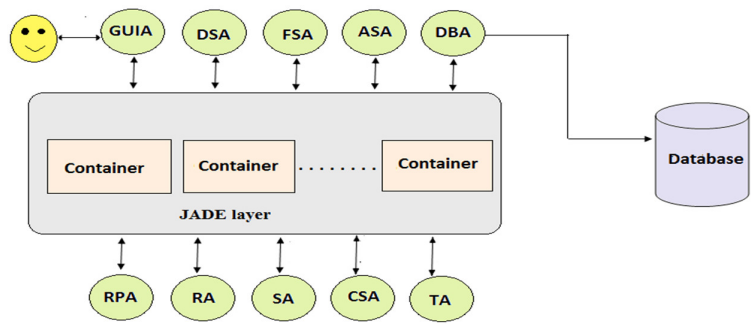

Figure 1: University IMS based on Multi-Agent architecture.

They are autonomous and can collect information from the environment and can make decisions. Each agent is named according to its functionality. Figure-1 shows the architecture of the system, consisting of functional agents.
All agents are connected to JADE layer which provides services for interaction and coordination between agents within the system and the user. Table-1 presents a short description of each agent.

**Table 1:** Agents present in University IMS Architecture

| Agent | Description |
|---|---|
| Graphical User Interface Agent (GUIA) | It provides the graphical interface for direct interaction of user with the system. It is a common way of retrieving, storing and sharing information in the system. Users get access to the database to store latest knowledge for the visitors. |
| Admission Schedule Agent (ASA) | It maintains information about admissions in all the academic programs of the department of computer science. Admission criteria, opening and closing dates, new offered programs etc. |
| Date Sheet Agent (DSA) | It manages date, time, venue, program, semester, department details of each examination paper. |
| Fee Structure Agent (FSA) | It manages fee structure of each semester. There are multiple study programs with each program consisting of multiple semesters, and with each semester having different fee structure. It defines new fee structure for the new offered courses. |
| Teacher Agent (TA) | Teacher creates an account by filling a simple registration form. This registration information would be managed by TA agent. |
| Student Agent (SA) | Student creates an account by filling a simple registration form. Information about students, their names, id, discipline all are managed by this agent. |
| Report Agent (RPA) | It creates statistical reports comprising of the number of students graduating each year, number of student admissions, student attendances, teacher to student ratio, computer laboratory to student ratio, reports consisting of graphs. |
| Result Agent (RA) | The results of each class would be calculated and managed by the Result agent. |
| Class Schedule Agent (CSA) | Class Schedule of each program is managed such that there is no conflict in timings of different courses of the same class. |
| Orchestrator Agent (OA) | It communicates directly with the database. All the tasks of information management (i.e. adding new information, updating data, calculating results, and generating reports) are handled by OA. |

Table 2: University IMS in terms of belief, desire and intention

| Agent | Description |
|---|---|
| **Belief** | Belief is an environmental knowledge. To get goals (desires) the system needs the information of: Teachers (name, designation, contact no, email), Discipline (name, duration, semester), Programs (evening, morning), Students (name, discipline, program, roll no) |
| **Desire** | Desires are the way that we pass through belief to action. Our goal here is to provide correct and updated information to the users according to their needs; to keep all agents up-to-date; and |
| | generate reports when needed by acquiring information from multiple different agents. For the generation of reports (i.e. system intention), these desires are necessary: Calculate results of students, Create class schedule for each discipline, Create admission schedule for each program, Update fee structures of each discipline and each program. |
| **Intention** | Achievement of goals directly affects our intentions. Every agent should keep data updated and consistent. An agent place desires to achieve intentions. The intentions are to generate reports from calculated results, managed schedules (e.g. admission schedule, class schedule), and fee structures of each discipline. |

The planning of all desires needs environmental knowledge that includes basic information of students, teachers, disciplines, and programs. Each agent has to interact with orchestrator agent to get access to the database for information access. For this purpose agent send a request to orchestrator agent which sends query to the database.

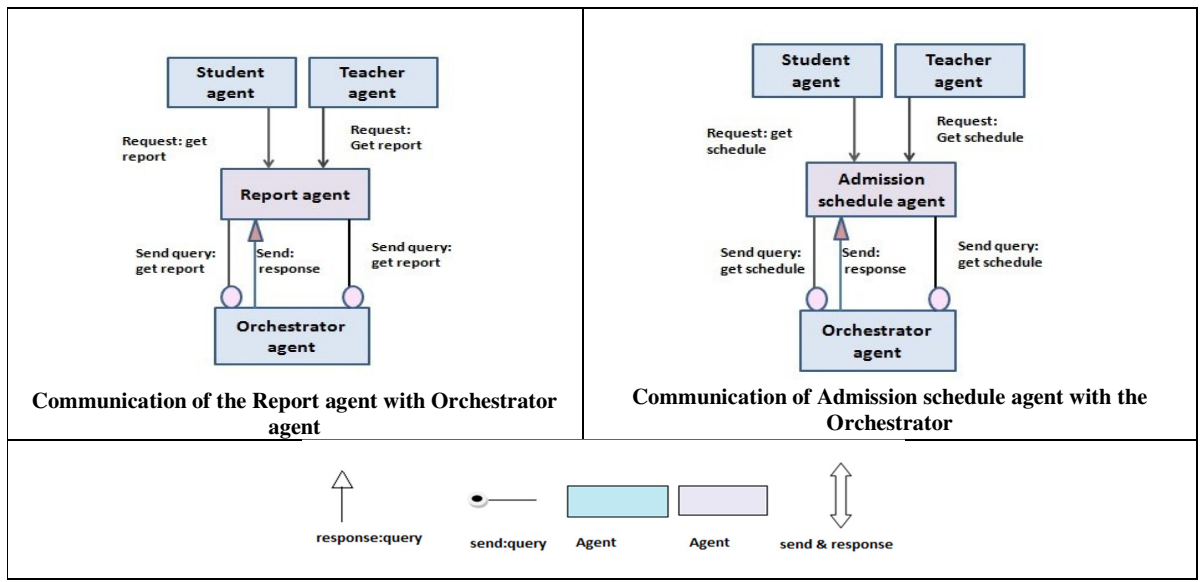

Figure 2: Representing Belief, Desire and Intention in the system

## 5.2. Verification (First-order predicate logic)

A student has to get registered in order to access the system. Each student will be registered and allotted a unique id. Duplicate registrations are avoided.

∃ **Student: (Request_GetRegister)**
$\Rightarrow$ ( ¬(Student.Has_id) $\Rightarrow$ (New.Student_id := SA.GenerateNew_id) )

Here SA is Student Agent that manages the student information. Teachers would be registered in the same way as students.

After getting registered, a student will request to get admission in a program. Each program has its unique program id. After getting request from the user, the system will check if student is already admitted or not. After this check the student gets admission.

∃ **Student: (Student.AdmissionRequest)** $\Rightarrow$( ¬(Student.Has_id) ∧ (Program.Has_id == !))
$\Rightarrow$ ( ¬(Student.Has_id) ∧ (Student.program_id := ASA.Program_id))

Here ASA is the Admission Schedule Agent that manages admission for each class in a program.

Class Schedule Agent in our system will manage the classes of each program. Classes would be managed in a way that the timings of two subjects of the same class would not conflict.

$\forall$**classes: (Add_Newclass)**
$\Rightarrow$ **(CSA.class_id := CSA.lastclass_id + 1)** $\wedge$ **( ¬(CSA.Subject1_timing == CSA.Subject2_Timing)**
$\forall$**classes: (Assign_Teacher)** $\Rightarrow$ **( ¬(CSA.Subject1_Teacher == CSA.Subject2_Teacher)**

Fee structure agent would manage the fee structure of each program. If during admission any new program would be offered the fee structure agent will sense and update the fee structure. Here e is the fee structure agent.

$\exists$**New_program: (Offer_New_Program)**
$\Rightarrow$ **(Program.new_id := FSA.new_id)** $\wedge$ **¬(Program.AlreadyAllotted_id)** $\wedge$ **(FSA.update_fee)**
$\qquad\Rightarrow$ **(FSA.NewProgram_fee:= Program.NewProgram_fee)**

Datesheet of each class would be managed by the datesheet agent. Midterm exams would be allowed at least after minimum numbers of lectures that are sixteen lecture and final term after minimum thirty two lectures.

$\exists$**class: (Manage_Datesheet)**
$\Rightarrow$**(¬(CSA.lectures_deliverd<minimum_lectures)** $\wedge$ **(CSA.lectures_deliverd>maximum_lectures))**
$\qquad\wedge$ **(set_subject_timing)** $\Rightarrow$ **(CSA.subject_timing := DSA.subject_timing)**
$\qquad\qquad\wedge$**¬(CSA.AlreadyAssigned_timing)** $\wedge$ **( ¬(CSA.Assigned_date**
$\qquad\qquad$**== DSA.AlreadyAssigned_date))**

Student result is calculated by the result agent. System will restrict that student can neither get more then maximum marks nor less then minimum marks and result can never be negative.

$\exists$**student: (Calculate_Result)**
$\qquad\Rightarrow$ **( ¬(RA.calculated_result<minimum_marks)** $\wedge$ **(RA.calculated_result>maximum_marks)**
$\qquad\qquad\wedge$ **(RA.calculated_result == negative))**

Reports are generated by the system. This report can be of student results, teachers to student ratio, number of students admitted in a year, number of teachers in a program.

$\forall$**users: (Generate_Report)** $\Rightarrow$ **(RPA.GenerateReport)**

### 5.3. Safety properties

Some properties of IMS are given below:

*Student registration property*

If any student got registered one time then he cannot be registered again
$\exists$ *student  (( student_id == registerd_id)  $\Rightarrow$ display ("Student Already Registerd"))*
student_id is the students' Citizen National Identity Card no and registerd_id is the list of all ids that are already registered

*Scalability property*

The scalability property of accessing data and generating reports. The system is restricted to 1000 users that can access the system at one time. If more students send request they will receive a message "busy".
$\forall w$  *((userno>1000)  $\Rightarrow$ display("busy"))*
Here **w** denoted the user and **userno** denoted the number of user requests for system access. **If** userno will be greater **than** 1000 then system will display a message of busy.

*Access property*

Consider the access property of the system.

$\forall x \ ((\neg (dpt\_id == get\_csid) \Rightarrow display\ ("unauthorized\ access"))$

This property said candidate who try to get access to the system and they do not belong to Computer Science & IT department. The **dpt_id** will be provided by university to each student according to their departments. **get_csid** is the list of cs department ids'. System will chek the user against that id.

*Admission property*

Consider the property of duplicate admission request.

$\forall S \ (request\_admission\ (p\_id) \land (st\_id) == registered\ (p\_id) \land (st\_id)$
$\Rightarrow display\ ("duplicate\ admission\ request")\ )$

When a student request to get admission in the dpt he mentions p_id and st_id the system checks the student against that id. If these ids' already exist in the system it means student is already admitted in any program so the request of the student would not be proceeded. Here **p_id** is the program id and st_id is the student CNIC no.

*Fee structure property*

Consider the property of fee structure of newly added programs in admission schedule agent. It's a responsibility of Fee structure to keep sensing the changes in admission schedule agent and keep up to date fee structures.

$\forall P \ ((add\_newprogram(p) \Rightarrow update\_Fee\ (f))$

This property said that If any new program will be added in admission schedule, fee structure must be updated.

*Time conflict property*

Consider the property of timing of different classes. Class schedule agent performs the time management of different classes.

$\forall C \ (T(c1) == T(c2) \Rightarrow display\ ("same\ timing")\ )$

T is the timing and c1, c2 are classes, timing of two classes of a program must be different.

*Term property*

Consider the property of mid term and final term exams

$\forall C \ ( \neg(Lec\_deliverd(L) < min) \land ( \neg Lec\_deliverd(L) > max) \land ( \neg(Lec\_deliverd(L) == Negative))$
$min = 16 \ \ max = 32$

Datesheet of mid term and final term would be after minimum 16 lectures and final term would be after 32 lectures after first day of starting class.

*Date sheet conflict property*

Consider the property of datesheet timing.

$\forall C \ (Ex\_date\ (p1)(c1)(sub1) == Ex\_date\ (p1)(c1)(sub2) \Rightarrow display\ ("same\ date\ conflict"))$
$p1 = program \ \ c1 = class \ \ sub1, sub2 = subject.$

This property said while making the date sheets it must be concerned that two papers of the same class cannot be conducted on the same day.

*No Loss property*

Consider the property of no loss of data. It ensures there is no data loss during uploading, updating, and report generation.

$\forall x \ (update(x) \land \forall y( \neg(y=!)))$

This property said all required fields are updated and no field is empty in database. Here **x** denoted the fields that are to be updated.

*Calculation property*

Consider Safety Property of result agent, which calculate the results that will be accurate. Result marks of student cannot exceed from maximum marks and cannot decrease from minimum marks and cannot be negative ever.

$$\forall x \ ( \neg(calculated\_result(r)>max) \land ( \neg(calculated\_result(r)<min)) \land (\neg(calcuted\_result(r)==negative)))$$

Here max denoted the maximum marks and min denoted minimum marks.

*Report not null property*

Report Agent is the most sensitive agentive .To ensures correctness of the report agent must keep sensing the changing's in the environment and other agents. When information is going to be retrieved in the form of report it will never give null.

$$\forall x \ ( \neg(getReport(x)==!))$$

*Scalability property*

Consider scalability property of accessing data or generating reports. The system is restricted to 1000 users that can access the system at a time. If more students send request they will receive a message "busy".

$$\forall w \ ((userno> 1000) \Rightarrow display("busy"))$$

Here **w** denotes the users and **userno** denotes the number of user who is requesting for accessing the system. **If** userno will be greater **than** 1000 then system will display a message of busy.

## 6. CONCLUSION

This paper presents a BDI [Bratman, 1987] [Georgeff et al., 1994] based multi-agent IMS to access, update and share information with the objectives to formally specify the safety and liveness properties by using First-Order Predicate Logic. The system is composed of BDI agents that are autonomous. The proposed BDI architecture is composed of multiple agents and the task of each agent is defined. The safety properties are identified and specified by analyzing the system requirements.

In future there would be two axis of work. One is to compile and model our system properties in VDM [Jones, 1990]. Second is to propose a global as well as detailed architectural design of the system.

## REFERENCES

Bratman, Michael E. (1987). Intentions, Plans, and Practical Reason. Harvard University Press, Cambridge.

Burkhard, H. D. (1996), Abstract Goals in Multi-Agent Systems. In Twelfth European Conference on Artificial Intelligence (ECAI96), ed. W. Wahlster, New York: Wiley, pp. 524–528.

Busetta, P. Ronnquist, R. Hodgson, A. and Lucas. A.(1999) JACK Intelligent Agents: Components for intelligent agents in Java. AgentLinkNewsLetter. AOS Pty. Ltd.

Demazeau, Y. (2006). Multi-Agent Systems Methodology. In Second Franco-Mexican School on Cooperative and Distributed Systems (LAFMI 2003),

Evertsz, R., Ritter, F. F., Busseta, P. and Pedrotti, M. (2008). Realistic variation in a BDI based cognitive architecture. In proceedings of SimTecT.

Georgeff, M. P. and Lanasky, A. L. (1987). Reactive reasoning and planning. In Proceedings of the Sixth National Conference on Artificial Intelligence (AAAI-87), Seattle, WA, pp. 677-682.

Geylani, K. (2013) Model-driven development of multi-agent systems: a survey and evaluation. The Knowledge Engineering Review. Cambridge University Press. pp.1.

Hindriks, K. V., De Boer, F. S., der Hoek, W. V., and Meyer J-J., Ch. (1999), Agent Programming in 3APL, Autonomous Agents and Multi-Agent Systems, (2)4 pp. 357-401.

Howden N, Ronnquist R, Hodgson A, Lucas A (2001) JACK intelligent agents — summary of an agent infrastructure. In Proceedings of the 5th International Conference on Autonomous Agents, Montreal.

Jennings, N. and Wooldridge, M. (1996) Software agents, IEEE Review, pp. 17-20.

Jennings, N. R., and Wooldridge, M. (1994). Proceedings of the ECAI-94 Workshop on Agent Theories, Architectures, and Languages. Lecture Notes in Artificial Intelligence 890. New York: Springer-Verlag,

Jones, C. B. (1990). Systematic Software Development Using VDM, 2nd ed. Prentice-Hall International, Englewood Cliffs, NJ.

Rao, A. S., and Georgeff, M. P. (1995). BDI Agents: From Theory to Practice. Lesser, V. (Ed), In Proceedings of the First International Conference on Multi-Agent Systems (ICMAS- 95). Mass.: MIT Press, Cambridge, pp. 312–319.

Rao, A. S. (1996). AgentSpeak (L): BDI Agents Speak Out in a Logical Computable Language. In W. V. de Velde and J. W. Perram, editors, MAAMAW, volume 1038 of Lecture Notes in Computer Science, pages 42–55. Springer

Rens, G., Ferrein, A and Van der Poel, E. (1-3 June, 2009). BDI agent architecture for a POMDP planner. 9th International Symposium on Logical Formalization of Commonsense Reasoning: Commonsense, Toronto, Canada. pp. 6.

Zambonelli, F., Jennings, N. R., and Wooldridge, M. (2003). Developing Multiagent Systems: The Gaia Methodology. ACM Transactions on Software Engineering and Methodology, 12(3), pp. 317-370.